\documentstyle[12pt]{article}

\catcode`@=11
\def\un#1{\relax\ifmmode\@@underline#1\else
        $\@@underline{\hbox{#1}}$\relax\fi}
\catcode`@=12

\def\a{\alpha}
\def\b{\beta}

\def\d{\delta}

\def\g{\gamma}
\def\h{\eta}

\def\l{\lambda}

\def\p{\pi}

\def\L{\Lambda}

\def\bo{{\raise-.5ex\hbox{\large$\Box$}}}               % D'Alembertian
\def\pa{\partial}                                       % curly d
\def\TH{{\raise.2ex\hbox{$\displaystyle \bigodot$}\mskip-4.7mu \llap H \;}}
\def\face{{\raise.2ex\hbox{$\displaystyle \bigodot$}\mskip-2.2mu \llap {$\ddot
        \smile$}}}                                      % happy face
\def\sp#1{{}^{#1}}                              % superscript (unaligned)
\def\Hat#1{\widehat{#1}}                        % big hat
\def\leftrightarrowfill{$\mathsurround=0pt \mathord\leftarrow \mkern-6mu
        \cleaders\hbox{$\mkern-2mu \mathord- \mkern-2mu$}\hfill
        \mkern-6mu \mathord\rightarrow$}
\def\dvec#1{\vbox{\ialign{##\crcr
        \leftrightarrowfill\crcr\noalign{\kern-1pt\nointerlineskip}
        $\hfil\displaystyle{#1}\hfil$\crcr}}}           % <--> accent
\def\frac#1#2{{\textstyle{#1\over\vphantom2\smash{\raise.20ex
        \hbox{$\scriptstyle{#2}$}}}}}                   % fraction
\def\ha{\frac12}                                        % 1/2
\def\sfrac#1#2{{\vphantom1\smash{\lower.5ex\hbox{\small$#1$}}\over
        \vphantom1\smash{\raise.4ex\hbox{\small$#2$}}}} % alternate fraction
\def\bfrac#1#2{{\vphantom1\smash{\lower.5ex\hbox{$#1$}}\over
        \vphantom1\smash{\raise.3ex\hbox{$#2$}}}}       % "
\def\afrac#1#2{{\vphantom1\smash{\lower.5ex\hbox{$#1$}}\over#2}}    % "
\def\[{\lfloor{\hskip 0.35pt}\!\!\!\lceil}
\def\]{\rfloor{\hskip 0.35pt}\!\!\!\rceil}

\def\ud#1#2{^{#1}{}_{#2}}

\def\fracm#1#2{\hbox{\large{${\frac{{#1}}{{#2}}}$}}}
\def\half{{\fracm12}}
\def\ha{\half}
\def\tr{{\rm tr}}

\def\un{\underline}
\def\fracmm#1#2{{{#1}\over{#2}}}

\def\low#1{{\raise -3pt\hbox{${\hskip 0.75pt}\!_{#1}$}}}

\def\Hat#1{\widehat{#1}}

\newskip\humongous \humongous=0pt plus 1000pt minus 1000pt
\def\caja{\mathsurround=0pt}
\def\eqalign#1{\,\vcenter{\openup2\jot \caja
        \ialign{\strut \hfil$\displaystyle{##}$&$
        \displaystyle{{}##}$\hfil\crcr#1\crcr}}\,}
\newif\ifdtup

\def\ref#1{$\sp{#1)}$}

\def\pl#1#2#3{Phys.~Lett.~{\bf {#1}B} (19{#2}) #3}
\def\np#1#2#3{Nucl.~Phys.~{\bf B{#1}} (19{#2}) #3}

\parskip=\medskipamount                 % space between paragraphs (LaTeX)
\thicklines                         % thick straight lines for pictures (LaTeX)

\begin{document}

% =========================== title page ==================================

\thispagestyle{empty}               % no heading or foot on title page (LaTeX)

\def\border{                                            % border
        \setlength{\unitlength}{1mm}
        \newcount\xco
        \newcount\yco
        \xco=-24
        \yco=12
        \begin{picture}(140,0)
        \put(-20,11){\tiny Institut f\"ur Theoretische Physik Universit\"at
Hannover~~ Institut f\"ur Theoretische Physik Universit\"at Hannover~~
Institut f\"ur Theoretische Physik Hannover}
        \put(-20,-241.5){\tiny Institut f\"ur Theoretische Physik Universit\"at
Hannover~~ Institut f\"ur Theoretische Physik Universit\"at Hannover~~
Institut f\"ur Theoretische Physik Hannover}
        \end{picture}
        \par\vskip-8mm}

\def\headpic{                                           % UH heading
        \indent
        \setlength{\unitlength}{.8mm}
        \thinlines
        \par
        \begin{picture}(29,16)
        \put(75,16){\line(1,0){4}}
        \put(80,16){\line(1,0){4}}
        \put(85,16){\line(1,0){4}}
        \put(92,16){\line(1,0){4}}

        \put(85,0){\line(1,0){4}}
        \put(89,8){\line(1,0){3}}
        \put(92,0){\line(1,0){4}}
        \put(85,0){\line(0,1){16}}
        \put(96,0){\line(0,1){16}}
        \put(79,0){\line(0,1){16}}
        \put(80,0){\line(0,1){16}}
        \put(89,0){\line(0,1){16}}
        \put(92,0){\line(0,1){16}}
        \put(79,16){\oval(8,32)[bl]}
        \put(80,16){\oval(8,32)[br]}

        \end{picture}
        \par\vskip-6.5mm
        \thicklines}

\border\headpic {\hbox to\hsize{
\vbox{\noindent ITP--UH--11/94 \\ hep-th/9409007
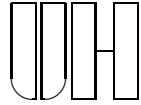
\hfill September 1994}}}

\noindent
\vskip1.3cm
\begin{center}

{\Large\bf BRST Charge for the Orthogonal Series of
           Bershadsky-Knizhnik Quasi-Superconformal
                         Algebras                  }
\footnote{Supported in part by the `Deutsche Forschungsgemeinschaft' and the
NATO grant CRG 930789}\\
\vglue.3in

Sergei V. Ketov \footnote{On leave of absence from:
High Current Electronics Institute of the Russian Academy of Sciences,
Siberian Branch, Akademichesky~4, Tomsk 634055, Russia}

{\it Institut f\"ur Theoretische Physik, Universit\"at Hannover}\\
{\it Appelstra\ss{}e 2, 30167 Hannover, Germany}\\
{\sl ketov@kastor.itp.uni-hannover.de}
\end{center}
\vglue.3in

\begin{center}
{\Large\bf Abstract}
\end{center}
\vglue.1in
\begin{quote}
The quantum BRST charges for Bershadsky-Knizhnik orthogonal
quasi-superconformal algebras are constructed. These two-dimensional
superalgebras have the $N$-extended non-linearly realised supersymmetry and
the $SO(N)$ internal symmetry. The BRST charge nilpotency conditions are shown
to have a unique solution at $N>2$, namely, $N=4$ and $k=-2$, where $k$ is
central extension parameter of the Ka\v{c}-Moody subalgebra. We argue about
the existence of a new string theory with the non-linearly realised $N=4$
world-sheet supersymmetry and negative `critical dimension'.
\end{quote}

\newpage
\hfuzz=10pt

{\bf 1}. Any known critical $N$-extended fermionic string theory with $N\leq 4$
 world-sheet supersymmetries is based on a two-dimensional (2d) linear
$N$-extended superconformal algebra which is gauged \cite{aba}. When a number
of world-sheet supersymmetries exceeds two, there are more opportunities to
build up new string theories, namely, by utilyzing 2d non-linear {\it
quasi-superconformal algebras} (QSCAs) which are known to exist for an
arbitrary $N>2$. The QSCAs can be considered on equal footing with the $W$
algebras {\it without}, however, having currents of spin higher than two. In
the past,  only one string theory for $N>2$ was actually constructed
 by gauging the `small' linear $N=4$ SCA with $SU(2)$ internal
symmetry~\cite{aba,pn}. Still, it is of interest to know how many
 different $N=4$ string theories exist at
all. Any $N=4$ string constraints are going to be very strong, so that their
explicit realisation should always imply  non-trivial interplay between
geometry, conformal invariance and extedned supersymmetry. The $N=4$ strings
are also going to be relevant in the search for the `universal string theory'
\cite{bvu}. In addition, strings with $N=4$ supersymmetry are expected to have
deep connections with integrable models \cite{ket,book}, so that we believe
they are worthy to be studied.

The full classification of QSCAs is known due to by Fradkin and Linetsky
\cite{fl}. In particular,  the $osp(N|2;{\bf R})$ and $su(1,1|N)$ series
of QSCAs with $\Hat{SO(N)}$ and $\Hat{U(N)}$ {\it Ka\v{c}-Moody} (KM)
symmetries, respectively, were discovered before by Knizhnik \cite{kn} and
Bershadsky \cite{be}. It has been known for some time that the {\it unitary}
 series of Bershadsky-Knizhnik QSCAs does not admit nilpotent quantum BRST
charges for any $N>2$ \cite{ssvn}, so that we are going to
concentrate on the orthogonal series of QSCAs having the $SO(N)$
internal symmetry.~\footnote{A construction of quantum BRST charges for the
orthogonal series of QSCAs was also briefly discussed in ref.~\cite{ssvn}, but
the results presented there are, however, incomplete.}
\vglue.2in

{\bf 2}. The current contents of the 2d Bershadsky-Knizhnik orthogonal QSCA
\cite{kn,be} is given by the holomorphic fields $T(z)$, $G^i(z)$ and
$J^a(z)$, all having the standard mode expansions
$$\eqalign{
T(z)= & \sum_n L_n z^{-n-2}~,\cr
G^i(z)= & \sum_r G^i_r z^{-r-3/2}~,\cr
J^a(z)= & \sum_n J^a_n z^{-n-1}~,\cr}\eqno(1)$$
and (conformal) dimensions $2$, $3/2$ and $1$, respectively. The
supercurrents $G^i$, $i=1,\ldots,N$, are defined in the fundamental
representation of the internal symmetry group $SO(N)$ generated by
the zero modes $J^a_0$, $a=1,\ldots,{\frac 12}N(N-1)$, in the adjoint
representation.

Most of the {\it operator product expansions} (OPEs) defining an
orthogonal Bershadsky-Knizhnik QSCA take the standard linear form,
 {\it viz.}
$$\eqalign{
T(z)T(w)~\sim~ & \fracmm{c/2}{(z-w)^4} + \fracmm{2T(w)}{(z-w)^2}
+ \fracmm{\pa T(w)}{z-w}~,\cr
T(z)G^i(w)~\sim~ & \fracmm{{\frac 32}G^i(w)}{(z-w)^2} +
 \fracmm{\pa G^i(w)}{z-w}~,\cr
T(z)J^a(w)~\sim~ & \fracmm{J^a(w)}{(z-w)^2} + \fracmm{\pa J^a(w)}{z-w}~,\cr
J^a(z)G^i(w)~\sim~ & \fracmm{(t^a)^{ij}G^j(w)}{z-w}~,\cr
J^a(z)J^b(w)~\sim~ & \fracmm{f^{abc}J^c(w)}{z-w}
+ \fracmm{-k\d^{ab}}{(z-w)^2}~,\cr}\eqno(2)$$
where $k$ is an arbitrary `level' of the KM subalgebra, $f^{abc}$ are $SO(N)$
structure constants, and $(t^a)^{ij}$ are generators of $SO(N)$ in the
fundamental (vector) representation,
$$\eqalign{
\[t^a,t^b\]=f^{abc}t^c,\quad & \quad  f^{abc}f^{abd}=2(N-2)\d^{cd}~,\cr
\tr(t^at^b)=-2\d^{ab}~,\quad & \quad
 (t^a)^{ij}(t^a)^{kl}=\d^{ik}\d^{jl}-\d^{il}\d^{jk}~.\cr}\eqno(3)$$

On symmetry and dimensional reasons, the only (non-linear) OPE defining the
supersymmetry piece of QSCA can be of the form
$$\eqalign{
G^i(z)G^j(w) ~\sim~ & a_1\fracmm{\d^{ij}}{(z-w)^3} + a_2\fracmm{(t^a)^{ij}
J^a(w)}{(z-w)^2} +\fracmm{1}{z-w}\left[ 2\d^{ij}T(w)
+ \ha a_2(t^a)^{ij}\pa J^a(w)\right]\cr
& +\fracmm{1}{z-w}\left[ a_3\left(t^{(a}t^{b)}\right)^{ij}
+a_4\d^{ab}\d^{ij}\right]:J^aJ^b:(w)~,\cr}\eqno(4)$$
where $a_1,\,a_2,\,a_3$ and $a_4$ are parameters to be determined by
solving the Jacobi identity, and the normal ordering is defined by
$$:J^aJ^b:(w)=\lim_{z\to w}\left[J^{(a}(z)J^{b)}(w) +
 \fracmm{k\d^{ab}}{(z-w)^2}\right]~.\eqno(5)$$
Indices in brackets mean symmetrization with unit weight,
e.g. $t^{(a}t^{b)}\equiv {\frac 12}(t^at^b+t^bt^a)$. Eq.~(4) can be
considered as the general {\it ansatz} for supersymmetry algebra.

Demanding consistency of the whole algebra determines the parameters
\cite{kn,be}:
$$
a_1=\fracmm{k(N+2k-4)}{N+k-3}~,\quad  a_2= \fracmm{N+2k-4}{N+k-3}~,\quad
a_3=  a_4 = \fracmm{1}{N+k-3}~, \eqno(6)$$
while the Virasoro central charge of this QSCA is also quantized as
\cite{kn,be}:
$$ c = \fracmm{k(N^2+6k-10)}{2(N+k-3)}~.\eqno(7)$$
The KM parameter $k$ remains arbitrary in this construction.

In case of the $N=2$ QSCA, the non-linearity actually disappears  and the
algebra becomes the $N=2$ linear SCA, since the total coefficient in front of
the sum of two last terms in the second line of eq.~(4) vanishes ($i^2+1=0$)
after substituting $U(1)\cong SO(2)$ and the last eq.~(6). Therefore, the
non-linear structure of Bershadsky-Knizhnik QSCAs only appears for $N\geq 3$.
\vglue.2in

{\bf 3}.  Despite of the apparent non-linearity of QSCAs, their
quantum BRST charges should be in correspondence with their classical BRST
charges, up to renormalisation. The classical procedure is known for an
arbitrary algebra of first-class constraints~\cite{ff}. It was already
used to obtain the quantum BRST charge for the non-linear $W_3$ algebra
\cite{tm}, and later generalised to any quadratically non-linear
$W$-type algebra in ref.~\cite{ssvn}.

Consider a set of bosonic generators $B_i$ and fermionic generators $F_{\a}$,
which satisfy a graded non-linear algebra of the form
$$\eqalign{
\{B_i,B_j\}_{\rm P.B.}=~&~f_{ij}{}^kB_k~,\cr
\{B_i,F_{\a}\}_{\rm P.B.}=~&~f_{i\a}{}^{\b}F_{\b}~,\cr
\{F_{\a},F_{\b}\}_{\rm P.B.}=~&~f_{\a\b}{}^{i}B_{i}+\L_{\a\b}{}^{ij}B_iB_j~,
\cr}\eqno(8)$$
in terms of the graded Poisson (or Dirac) brackets, with some 3-point and
4-point `structure constants', $f_{ij}{}^k$, $f_{i\a}{}^{\b}$, $f_{\a\b}{}^{i}$
 and $\L_{\a\b}{}^{ij}$, respectively, which have to be ordinary numbers. The
symmetry properties of these constants with respect to exchanging their indices
 obviously follow from their definition by eq.~(8), and they are assumed below.
When using the unified index notation, $A\equiv(i,\a),\ldots~$, the Jacobi
identities for the classical graded algebra of eq.~(8) take the form
$$\eqalign{
f_{[AB}{}^{D}f_{C\}D}{}^E~=~&~0~,\cr
\L_{[AB}{}^{DE}f_{C\}D}{}^F +\L_{[AB}{}^{DF}f_{C\}}{}^E
+f_{[AB}{}^D\L_{C\}D}{}^{EF}~=~&~0~.\cr}\eqno(9)$$
As is clear from eq.~(9), $f_{AB}{}^C$ are to be the structure constants of
a graded Lie algebra.~\footnote{We assume that all symmetry operations with
unified indices also have to be understood in the graded sense. In particular,
a graded `antisymmetrisation' of indices with unit weight (denoted by mixed
brackets $\[\;\}$ here) actually means the antisymmetrisation for
bosonic-bosonic or bosonic-fermionic index pairs, but the symmetrisation for
indices which are both fermionic.}
According to the classical BRST procedure,~\footnote{See, e.g., ref.~\cite{hen}
for a review.} one introduces an anticommuting ghost-antighost pair
$(c^m,b_m)$ for each of the bosonic generators $B_m$, and the commuting
ghost-antighost pair $(\g^{\a},\b_{\a})$ for each of the fermionic generators
$F_{\a}$. The ghosts satisfy (graded) bracket relations
$$\eqalign{
\{c^m,c^n\}_{\rm P.B.}=\{b_m,b_n\}_{\rm P.B.}=0~,~&~\{c^m,b_n\}_{\rm P.B.}=
\d\ud{m}{n}~,\cr
\{\g^{\a},\g^{\b}\}_{\rm P.B.}=\{\b_{\a},\b_{\b}\}_{\rm P.B.}=0~,~&~
\{\g^{\a},\b_{\b}\}_{\rm P.B.}=\d\ud{\a}{\b}~.\cr}\eqno(10)$$

Additional ghosts for the composite generators $B^iB^j$ are not needed since
invariance of the classical theory under $B^i$ already implies
invariance under $B^iB^j$ \cite{ff}.

The classial BRST charge $Q$ is  given by \cite{ff}
$$\eqalign{
Q = ~&~ c^nB_n + \g^{\a}F_{\a} +\ha f_{ij}{}^kb_kc^jc^i
+ f_{i\a}{}^{\b}\b_{\b}\g^{\a}c^i -\ha f_{\a\b}{}^nb_n\g^{\b}\g^{\a} \cr
&  -\ha \L_{\a\b}{}^{ij}B_ib_j\g^{a}\g^{\b}
-\fracm{1}{24}\L_{\a\b}{}^{ij}\L_{\g\d}{}^{kl}f_{ik}{}^{m}
b_jb_lb_m\g^{\a}\g^{\b}\g^{\g}\g^{\d}~.\cr}\eqno(11)$$
Compared to the standard expression for the linear algebras $(\L=0)$, the
BRST charge of eq.~(11) has the additional $3$-(anti)ghost terms, dependent on
the initial bosonic generators $B_i$, and the $7$-(anti)ghost terms as well.
It is easy to check that the classical `master equation'
$$\{Q,Q\}_{\rm P.B.}=0\eqno(12)$$
follows from eq.~(9) and the related identity
$$\L_{ \{\a\b}{}^{ij}\L_{\g\d\} }{}^{kl}f_{ik}{}^{m}=
\L_{ \{\a\b}{}^{i[j}\L_{\g\d\} }{}^{|k|l}f_{ik}{}^{m]}~.\eqno(13)$$

The classical BRST charge (11) may serve as the starting point in a
construction of quantum BRST charge $Q_{\rm BRST}$ for the corresponding
 graded non-linear quantum algebra. Since we are actually interested in
 quantum QSCAs, we can assume that all operators are just currents, with a
holomorphic dependence on $z$ or, equivalently, with an additional
affine index (see eq.~(1), for example). In particular, in eq.~(10) one should
replace the (graded) Poisson brackets by (anti)commutators. In addition, in
quantum theory, one must take into account central extensions and the normal
ordering needed for defining products of bosonic generators.
This results in the quantum (anti)commutation relations
$$\eqalign{
\[B_i,B_j\]=~&~f_{ij}{}^kB_k +h_{ij}Z~,\cr
\[B_i,F_{\a}\]=~&~f_{i\a}{}^{\b}F_{\b}~,\cr
\{F_{\a},F_{\b}\}=~&~h_{\a\b}Z+f_{\a\b}{}^{i}B_{i}+\L_{\a\b}{}^{ij}:B_iB_j:~,
\cr}\eqno(14)$$
where the central charge generator $Z$ commutes with all the other generators,
and the new constants $h_{ij}$ and $h_{\a\b}$ are supposed to be restricted by
 the Jacobi identities.  Although no general
procedure seems to exist, which would explain how to `renormalise' the
naively quantised normally-ordered charge $Q$ to a nilpotent quantum-mechanical
 operator $Q_{\rm BRST}$, the answer is known for a particular class of
quantum algebras of the $W$-type~\cite{ssvn}. Similarly to the quantum $W_3$
algebra case considered in ref.~\cite{tm}, a non-trivial modification of
eq.~(11) in quantum theory essentially amounts to a {\it multiplicative}
renormalisation of the structure constants $f_{\a\b}{}^i$, namely
$$\eqalign{
Q_{\rm BRST} = ~&~ c^nB_n + \g^{\a}F_{\a} +\ha f_{ij}{}^k:b_kc^jc^i:
+ f_{i\a}{}^{\b}:\b_{\b}\g^{\a}:c^i
-\ha\h f_{\a\b}{}^nb_n\g^{\b}\g^{\a} \cr
&  -\ha \L_{\a\b}{}^{ij}B_ib_j\g^{a}\g^{\b}
-\fracm{1}{24}\L_{\a\b}{}^{ij}\L_{\g\d}{}^{kl}f_{ik}{}^{m}
b_jb_lb_m\g^{\a}\g^{\b}\g^{\g}\g^{\d}~.\cr}\eqno(15)$$
This {\it ansatz} for the quantum BRST operator introduces only one additional
renormalisation parameter $\h$ to be determined from the BRST charge
nilpotency condition. Since the central extension parameters of the quantum
non-linear algebra are severely restricted by the Jacobi identities, whereas
the quantum BRST charge nilpotency condition could lead to some more
restrictions on their values, this construction procedure could make them
overdetermined, in general. Therefore, the existence of a quantum BRST charge
is not guaranteed, and it is important to check consistency in each particular
 case.
\vglue.2in

{\bf 4}. For the BRST quantisation of Berschadsky-Knizhnik $SO(N)$-based QSCA
the following ghosts are needed:
\begin{itemize}
\item the conformal ghosts ($b,c$), an anticommuting pair of
world-sheet free fermions of conformal dimensions~($2,-1$), respectively;
\item the $N$-extended superconformal ghosts ($\b^i,\g^i$) of conformal
dimensions~($\frac32,-\frac12$), respectively, in the fundamental (vector)
representation of $SO(N)$;
\item the $SO(N)$ internal symmetry ghosts ($\tilde{b}^a,\tilde{c}^a$) of
conformal dimensions~($1,0$), respectively, in the adjoint representation of
$SO(N)$.
\end{itemize}

The reparametrisation ghosts
$$b(z)\ =\ \sum_{n\in{\bf Z}} b_n z^{-n-2}~,\qquad
c(z)\ =\ \sum_{n\in{\bf Z}} c_n z^{-n+1}~,\eqno(16)$$
have the following OPE and anticommutation relations:
$$b(z)\ c(w)\ \sim\ \fracmm{1}{z-w}~,
\qquad \{c_m,b_n\}\ =\ \d_{m+n,0}~.\eqno(17)$$

The superconformal ghosts
$$\b^i(z)\ =\ \sum_{r\in{\bf Z}(+1/2)}\b^i_r z^{-r-3/2}~,\qquad
\g^i(z)\ =\ \sum_{r\in{\bf Z}(+1/2)}\g^i_r z^{-r+1/2}~,\eqno(18)$$
satisfy
$$\b^i(z)\ \g^j(w)\ \sim\ \fracmm{-\d^{ij}}{z-w}~, \qquad
\[\g^i_r,\b^j_s\]\ =\ \d_{r+s,0}~.\eqno(19)$$
An integer or half-integer moding of these generators corresponds to the usual
distinction between the Ramond- and Neveu-Schwarz--type sectors.

Finally, the fermionic $SO(N)$ ghosts
$$\tilde{b}^a(z)\ =\ \sum_{n\in{\bf Z}}\tilde{b}^a_n z^{-n-1}~,\qquad
\tilde{c}^a(z)\ =\ \sum_{n\in{\bf Z}} \tilde{c}^a_n z^{-n}~,\eqno(20)$$
have
$$\tilde{b}^a(z)\ \tilde{c}^a(w)\ \sim\ \fracmm{\d^{ab}}{z-w}~,\qquad
\{\tilde{c}^a_m,\tilde{b}^b_n\}\ =\ \d^{ab}\d_{m+n,0}~.\eqno(21)$$

Eq.~(15) provides us with the reasonable ansatz for the {\it quantum} BRST
charge,~\footnote{The normal ordering is implicit.}
$$\eqalign{
Q_{\rm BRST}=~&~ c_{-n}L_n + \g^i_{-r}G^i_r + \tilde{c}^a_{-n}J^a_n
-\ha (m-n)c_{-m}c_{-n}b_{m+n} + nc_{-m}\tilde{c}^a_{-n}\tilde{b}^a_{m+n}\cr
& +\left(\fracm{m}{2}-r\right)c_{-m}\b^i_{m+r}\g^i_{-r}
-b_{r+s}\g^i_{-r}\g^i_{-s} - \tilde{c}^a_{-m}\b^i_{m+r}(t^a)^{ij}\g^j_{-r}\cr
& + \h a_2(r-s)\tilde{b}^a_{r+s}\g^i_{-r}(t^a)^{ij}\g^j_{-s}
-\ha f^{abc}\tilde{c}^a_{-m}\tilde{c}^b_{-n}\tilde{b}^c_{m+n} \cr
& -\fracmm{1}{2}a_4\left[\left(t^{(a}t^{b)}\right)^{ij}+\d^{ab}\d^{ij}\right]
J^a_{r+s+m}\tilde{b}^b_{-m}\g^i_{-r}\g^j_{-s}
 -\fracmm{1}{24}a^2_4\left[\left(t^{(a}t^{b)}\right)^{ij}+\d^{ab}\d^{ij}\right]
\cr
&\times \left[\left(t^{(c}t^{d)}\right)^{kl}+\d^{cd}\d^{kl}\right]
 f^{ace}\d_{m+n+p,r+s+t+u}\tilde{b}^b_m\tilde{b}^d_n\tilde{b}^e_p
 \g^i_{-r}\g^j_{-s}\g^k_{-t}\g^l_{-u}~,\cr}\eqno(22)$$
where a quantum renormalisation parameter $\h$ has been introduced. Its value
is going to be fixed by the BRST charge nilpotency conditions. The coefficients
$a_2$ and $a_4$ have already been fixed by eq.~(6).

We find always useful to represent a quantum BRST charge as
$$Q_{\rm BRST}=\oint_0 \fracmm{dz}{2\p i}\,j_{\rm BRST}(z)~,\eqno(23)$$
where the BRST current $j_{\rm BRST}(z)$ is defined {\it modulo} total
derivative.~\footnote{The total derivative can be fixed by requring the
$j_{\rm BRST}(z)$ to transform as a primary field.} In particular, the BRST
current $j_{\rm BRST}(z)$ corresponding to the BRST charge of eq.~(22) is
given by
$$\eqalign{
j_{\rm BRST}(z)=~&~ cT +\g^i G^i + \tilde{c}^a J^a +bc\pa c
-c\tilde{b}^a\pa\tilde{c}^a
 -\ha c\g^i\pa\b^i-\fracm{3}{2}c\b^i\pa\g^i
-b\g^i\g^i\cr
& -\h a_2
\tilde{b}^a (t^a)^{ij}\left(\g^i\pa\g^j-\g^j\pa\g^i\right)
-\tilde{c}^a(t^a)^{ij}\b^i\g^j
 -\ha f^{abc}\tilde{c}^a\tilde{c}^b\tilde{b}^c\cr
& -\ha a_4
\left[\left(t^{(a}t^{b)}\right)^{ij}+\d^{ab}\d^{ij}\right]
J^a\tilde{b}^b\g^i\g^j  -\fracmm{1}{24} a^2_4
\left[\left(t^{(a}t^{b)}\right)^{ij}+\d^{ab}\d^{ij}\right]\cr
& \times \left[\left(t^{(c}t^{d)}\right)^{kl}+\d^{cd}\d^{kl}\right]f^{ace}
\tilde{b}^b\tilde{b}^d\tilde{b}^e\g^i\g^j\g^k\g^l~.\cr}\eqno(24)$$

The most tedious part of calculational handwork in computing $Q^2_{\rm BRST}$
can be avoided when using either the Mathematica Package for computing OPEs
\cite{th} or some of the general results in ref.~\cite{ssvn}. In particular,
as was shown in ref.~\cite{ssvn}, quantum renormalisation of the $3$-point
structure constants in the quantum BRST charge should be {\it multiplicative},
whereas the non-linearity $4$-point `structure constants' should {\it not} be
renormalised at all --- the facts already used in the BRST charge ansatz above.
 Most importantly, among the contributions to the $Q_{\rm BRST}^2$, only the
terms {\it quadratic} in the ghosts are relevant. Their vanishing imposes
 the constraints on the central extension coefficients of the QSCA and
simultaneously determines the renormalization parameter $\h$. The details
can be found in the appendices of ref.~\cite{ssvn}. The same conclusion comes
as a result of straightforward calculation on computer. Therefore, finding out
the nilpotency conditions amounts to calculating only a few terms `by hands',
namely, those which are quadratic in the ghosts. This makes the whole
calculation as simple as that in ordinary string theories based on linear SCAs
 \cite{book}.

The $2$-ghost terms in the $Q_{\rm BRST}^2$ arise from single contractions of
the first three linear (in the ghosts) terms of $Q_{\rm BRST}$ with themselves
and with the next cubic terms of eq.~(24), and from double contractions of
the latter among themselves. They result in the pole contributions to
$j_{\rm BRST}(z)j_{\rm BRST}(w)$, proportional to $(z-w)^{-n}$
with $n=1,2,3,4$. All the residues have to vanish modulo total derivative. We
find
$$\eqalign{
j_{\rm BRST}(z)j_{\rm BRST}(w)~\sim~&
 \fracmm{c(z)c(w)}{2(z-w)^4}\left[c-N^2+12N-26\right]  \cr
{}~&~+\fracmm{\g^i(z)\g^i(w)}{(z-w)^3}\left[a_1 -\fracmm{ka_4}{2}(N-1)(N-2)
-4\h a_2(N-1) +2\right]  \cr
{}~&~+\fracmm{\tilde{c}^a(z)\tilde{c}^a(w)}{(z-w)^2}\left[-k-2(N-2)+2\right]
\cr
{}~&~+\fracmm{J^a(w)(t^a)^{ij}\g^i(w)\pa\g^j(w)}{z-w}\left[-4\h a_2
-4a_4\left(1-\fracmm{N}{2}\right)\right]+\ldots~,  \cr}\eqno(25)$$
where the dots stand for the other terms of higher order in (anti)ghosts, and
the coefficients $a_1$, $a_2$, $a_4$ and $c$ are given by eqs.~(6) and (7),
 respectively. Eq.~(25) immediately yields the BRST charge nilpotency
conditions:
$$\eqalign{
c_{\rm tot}~&\equiv c +c_{\rm gh}~=~\fracmm{k(N^2+6k-10)}{2(N+k-3)}-N^2+12N-26
=0~,\cr
s_{\rm tot}~&\equiv ~a_1 + (a_1)_{\rm gh}~=~\fracmm{k(N+2k-4)}{N+k-3}
 -\fracmm{k(N-1)(N-2)}{2(N+k-3)}\cr
{}~&~\qquad\qquad\qquad~~~~~ -\fracmm{4\h(N-1)}{N+k-3} +2 = 0~,\cr
k_{\rm tot}~&\equiv ~k + k_{\rm gh}~=~k+2N-6=0~,\cr
&~\qquad\fracmm{ \h(N+2k-4)}{N+k-3}-\fracmm{N-2}{2(N+k-3)}~ = 0~.\cr}
\eqno(26)$$
The first line of eq.~(26) just means the vanishing total central charge, where
the value of $c_{\rm gh}$ is dictated by the standard formula of conformal
field theory \cite{book}
$$\eqalign{
c_{\rm gh}& = 2\sum_{\l} n_{\l}(-1)^{2\l+1}\left(6\l^2-6\l+1\right)\cr
& = 1\times (-26) + N\times (+11) + \ha N(N-1)\times (-2) = -26 +12N - N^2~,
\cr}\eqno(27)$$
$\l$ is conformal dimension and $n_{\l}$ is a number of the conjugated
ghost pairs: $\l=2,3/2,1$ and $n_{\l}=1,N,\ha N(N-1)$, respectively. The zero
central charge condition alone has two solutions,
$$k=6-2N~,\qquad {\rm and}\qquad 6k=N^2-12N+26~,\eqno(28)$$
but only the first of them is compatible with the third equation (26).

Central extensions (anomalies) of the ghost-extended QSCA need not form a
linear supermultiplet, and they actually do not. Therefore, the vanishing
central charge alone does not imply the other equations (26) to be
automatically satisfied, unlike in the linear case. The last equation (26)
just
determines the renormalisation parameter $\h$. Finally, the second equation
(26) can be interpreted as the vanishing total supersymmetric anomaly. Since
the supersymmetry is non-linearly realised, this anomaly does not have to
vanish as a consequence of the other equations (26), but restricts $N$ as the
only remaining parameter. Substituting
$$k=6-2N~,\qquad {\rm and }\qquad \h=\fracmm{N-2}{2(8-3N)}~,\eqno(29)$$
into the second equation (26), we find
$$6(N-3)+\fracmm{(N-1)(N-2)(N-5)}{N-3}=0~,\eqno(30)$$
which has only one solution, $N=4$. Therefore, though the system of four
equations (26) for only three parameters $\h$, $k$ and $N$ is clearly
overdetermined (while $N$ is a positive integer!), there is still the only
solution, namely
$$N=4~,\qquad k=-2~,\qquad \h=-\fracmm{1}{4}~.\eqno(31)$$
\vglue.2in

{\bf 5}. In our letter we constructed the quantum BRST charges for the
orthogonal  series of Bershadsky-Knizhnik non-linear QSCAs. The BRST charge
nilpotency conditions cannot be met, unless  $N=4$ and $k=-2$. This is
apparently in line with the analogous fact \cite{ssvn} that the BRST
quantisation breaks down for all unitary $U(N)$-based Bershadsky-Knizhnik
QSCAs of $N\geq 3$, since their BRST charge nilpotency conditions are always
in conflict with the Jacobi identities.~\footnote{The $U(N)$-based
Bershadsky-Knizhnik QSCAs do not admit unitary representations for $N\geq 3$
also~\cite{sch3}.}

The existence of the nilpotent quantum BRST charge for the non-linear
$SO(4)$-based Bershadsky-Knizhnik QSCA  implies the existence of a new
$W$-type string theory with the non-linearly realised $N=4$ world-sheet
 supersymmetry. In quantum theory, the vanishing currents can be
 interpreted as operator constraints on physical states. By interpreting
zero modes of  scalar fields in the matter QSCA realisations as spacetime
coordinates, one arrives at a first-quantised description of string
oscillations. Unfortunately, gauging the local symmetries of the $SO(4)$-based
Bershadsky-Knizhnik QSCA results in the positive total ghost central charge
contribution, $c_{\rm gh}=6$. In addition, the anomaly-free solution requires
$k=-2<0$. Therefore, there seems to be no way to build an anomaly-free string
theory when using only unitary representations of the $SO(4)$-based
Bershadsky-Knizhnik QSCA. When choosing a non-unitary representation of this
QSCA with $k=-2$, one can get the desired anomaly-free matter contribution,
$c_{\rm m}=-6$, but then a space-time interpretation together with a physical
significance of the construction, if any, become obscure. Despite of all this,
we believe that it is worthy to know how many string models, consistent from
the mathematical point of view, can be constructed.

\end{document}